\documentstyle[12pt,epsfig]{article}

\newcommand{\beq}{\begin{equation}}
\newcommand{\eeq}{\end{equation}}
\newcommand{\eq}[1]{(\ref{#1})}

\newcommand{\beqn}{\begin{eqnarray}}
\newcommand{\eeqn}{\end{eqnarray}}
\newcommand{\dst}{&\displaystyle}
\newcommand{\eps}{\mbox{$\varepsilon$}}
\newcommand{\T}{{\bf t}}

\newcommand{\p}{\mbox{${\bf p}$}}

\newcommand{\pch}{\mbox{$2\pi^+2\pi^-$}}
\newcommand{\pne}{\mbox{$\pi^+\pi^-2\pi^0$}}
\newcommand{\epm}{\mbox{$e^+e^-$}}

\begin{document}

\begin{titlepage}

\begin{center}
{\Large \bf Budker Institute of Nuclear Physics}
\end{center}

\vspace{1cm}

\begin{flushright}
{\bf Budker INP 99-56\\
June 09, 1999 }
\end{flushright}

\vspace{1.0cm}
\begin{center}{\Large \bf On the role of  $a_1(1260)$ meson in the
$\tau\to 4\pi\nu_{\tau}$ decay}\\
\end{center}
\vspace{1.0cm}

\begin{center}
A.E.Bondar, S.I.Eidelman, A.I.Milstein, N.I.Root
\\
  Budker Institute of Nuclear Physics, Novosibirsk, 630090, Russia
\end{center}

\begin{abstract}
We demonstrate that a simple model successfully describing
experimental data for the process
$\epm\to 4\pi$ can also qualitatively account for
the data of CLEOII and ALEPH
obtained recently for the decay $\tau\to 4\pi\nu_{\tau}$.
The model is based on the assumption of the $a_1(1260)\pi$ and
$\omega\pi$ dominance as intermediate states.
Our observation is in contrast with the claim by ARGUS 
that the $\rho$-meson signal in the
$\tau\to 4\pi\nu_\tau$ decay  can not be explained by the $a_1\pi$
intermediate state.
\end{abstract}

\end{titlepage}

\section{\bf Introduction}

At present hadron processes at low energies (below $J/\psi$ )
can not be satisfactorily described by QCD.
Therefore, in order to understand the phenomena in this
energy range it is necessary to apply different phenomenological models
and compare their predictions with the accessible experimental data.
Investigation of  exclusive channels in $\epm$ annihilation into
hadrons at low energies  and semihadronic decays of $\tau$ meson is
very attractive for this purpose.  It provides the important
information on the bound states of light quarks and elucidates the
role of different mechanisms in these processes.

Production of four pions is one of the dominant processes of
$\epm$ annihilation into hadrons in the energy range from 1.05 to 2.5 GeV.
Due to the conservation of vector current (CVC) the cross section of
this process is related to the probability of $\tau\to
4\pi\nu_{\tau}$ decay \cite{CVC1}. 
Therefore, all realistic models describing the
first process, should also be appropriate for the description 
of the other one.

One of the main difficulties in the study of
four pion production is caused by the existence of
different intermediate states via which the final state could be
produced, such as $\omega\pi$, $\rho \sigma$,
$a_{1}(1260) \pi$, $h_{1}(1170) \pi$, $\rho^{+} \rho^{-}$,
$a_{2}(1320) \pi$, $\pi(1300) \pi $.
The relative contributions of the mentioned processes can't
be obtained without the detailed analysis of the process dynamics.
Some information on the process $\epm\to 4\pi$ has been obtained 
in 
[2-8], where
the investigation of $\epm$ annihilation into hadrons
was restricted by measurements of the cross sections only.
However, the abundance of various possible mechanisms and their
complicated interference results in the necessity of simultaneous
analysis of two possible final states ($\pch$ and $\pne$) which requires
a general purpose detector capable of measuring
energies and angles of both charged and neutral particles.

In the energy range below 1.4 GeV the detector of this type CMD-2 is
operating at VEPP-2M collider in Novosibirsk \cite{cmddec}.
During a few last years  very large data samples have been obtained
which open qualitatively new  possibilities for the investigation of the
multihadronic production in $\epm$ annihilation.

Very recently, a model-dependent analysis was performed \cite{our}
of both possible channels in $\epm$ annihilation into four pions at
energies 1.05--1.38 GeV, using  the data collected with the CMD-2
detector. The discussion of previous works devoted to this theme
is also present in \cite{our}.

The detailed analysis of the process $\epm\to 4\pi$
unambiguously demonstrated that the main
contribution to the cross section  in the energy range
1.05 -- 1.38 GeV  , in addition to previously well-studied
$\omega\pi^0$  \cite{omegapi1,omegapi2},
is given by the $\rho\pi\pi$ intermediate state.
Moreover, the latter is completely
saturated by the $a_1\pi$ mechanism.
The contribution of other intermediate states was
estimated  to be less than 15 \% .

In this paper we use the assumption of the $a_1\pi$ dominance for
comparison of the available experimental data for   $\tau\to
4\pi\nu_{\tau}$ decay \cite{Argus,Aleph,Cleo1,Cleo2}
with the prediction of our model successfully
described the data of $\epm\to 4\pi$.

\section{Results and discussion}

The initial hadron state (referred to as $\tilde\rho$), which
decays into  four pions  has the $\rho$-meson quantum numbers.
Due to the conservation of vector current the probability $d\Gamma_1$
and $d\Gamma_2$ of $\tau^-\to\pi^-\pi^+\pi^-\pi^0\nu_\tau$ and
$\tau^-\to\pi^-\pi^0\pi^0\pi^0\nu_\tau$ decays, respectively, can be
written as
\beq\label{G1}
\frac{d\Gamma_i}{ds}=\frac{G^2\cos^2\theta}{96\pi^3 m_\tau^3}
(m_\tau^2+2s)(m_\tau^2-s)^2 R_{4\pi}\frac{dW_i}{W_1+W_2}
\eeq
where $R_{4\pi}$ is the ratio of the cross section $\epm\to 4\pi$ and
$\epm\to \mu^+\mu^-$, $dW_1$ and $dW_2$ are the probabilities of
$\tilde\rho^-$ decays into $\pi^-\pi^+\pi^-\pi^0$ and
$\pi^-\pi^0\pi^0\pi^0$, respectively. Let $dW_3$ and $dW_4$ be
the probabilities of $\tilde\rho^0$ decays into $\pi^+\pi^-\pi^+\pi^-$ and
$\pi^+\pi^-\pi^0\pi^0$, then due to the isospin invariance, we have
\beq
W_1=\frac{1}{2}W_3+W_4 \quad ,\quad W_2=\frac{1}{2}W_3\, .
\eeq
The explicit forms of the matrix elements, corresponding to $W_i$, are
presented in the Appendix. In order to get the predictions for $\tau$
decay we neglect the interference between $\omega\pi$ amplitude and
$a_1\pi$, and rewrite \eq{G1} in the following form:
\beqn\label{G11}
\frac{d\Gamma_1}{ds}&=&\frac{G^2\cos^2\theta}{96\pi^3 m_\tau^3}
(m_\tau^2+2s)(m_\tau^2-s)^2 \left[
 R_{\omega\pi}\frac{dW_{\omega}}{W_\omega} +
  R_{2\pi^+2\pi^-}\frac{dW_1}{W_3}\right] \nonumber\\
\frac{d\Gamma_2}{ds}&=&\frac{G^2\cos^2\theta}{96\pi^3 m_\tau^3}
(m_\tau^2+2s)(m_\tau^2-s)^2 R_{2\pi^+2\pi^-}
\frac{dW_2}{W_3} \ ,
\eeqn
where $dW_{\omega}$ is the probability of $\tilde\rho^-\to\omega\pi^-$
decay, $R_{\omega\pi}$ is the  ratio of the cross section
$\epm\to \omega\pi$ and $\epm\to \mu^+\mu^-$.

\begin{figure}[ptb]
\centering
\epsfig{figure=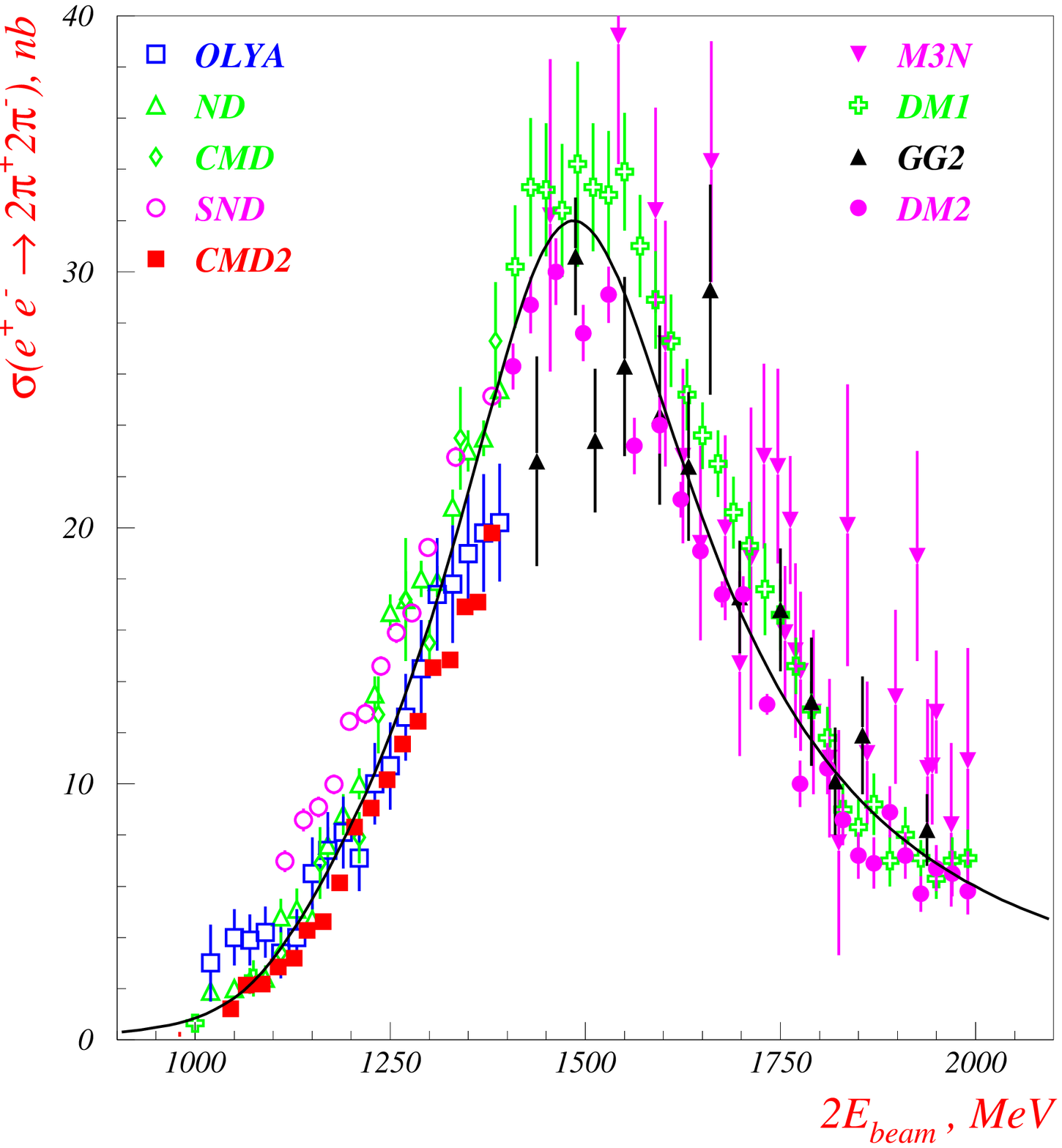,height=20cm}
\caption{
Energy dependence of the $e^+e^-\to 2\pi^+2\pi^-$ cross section
}
\label{xs4tau}
\end{figure}

\begin{figure}[ptb]
\centering
\epsfig{figure=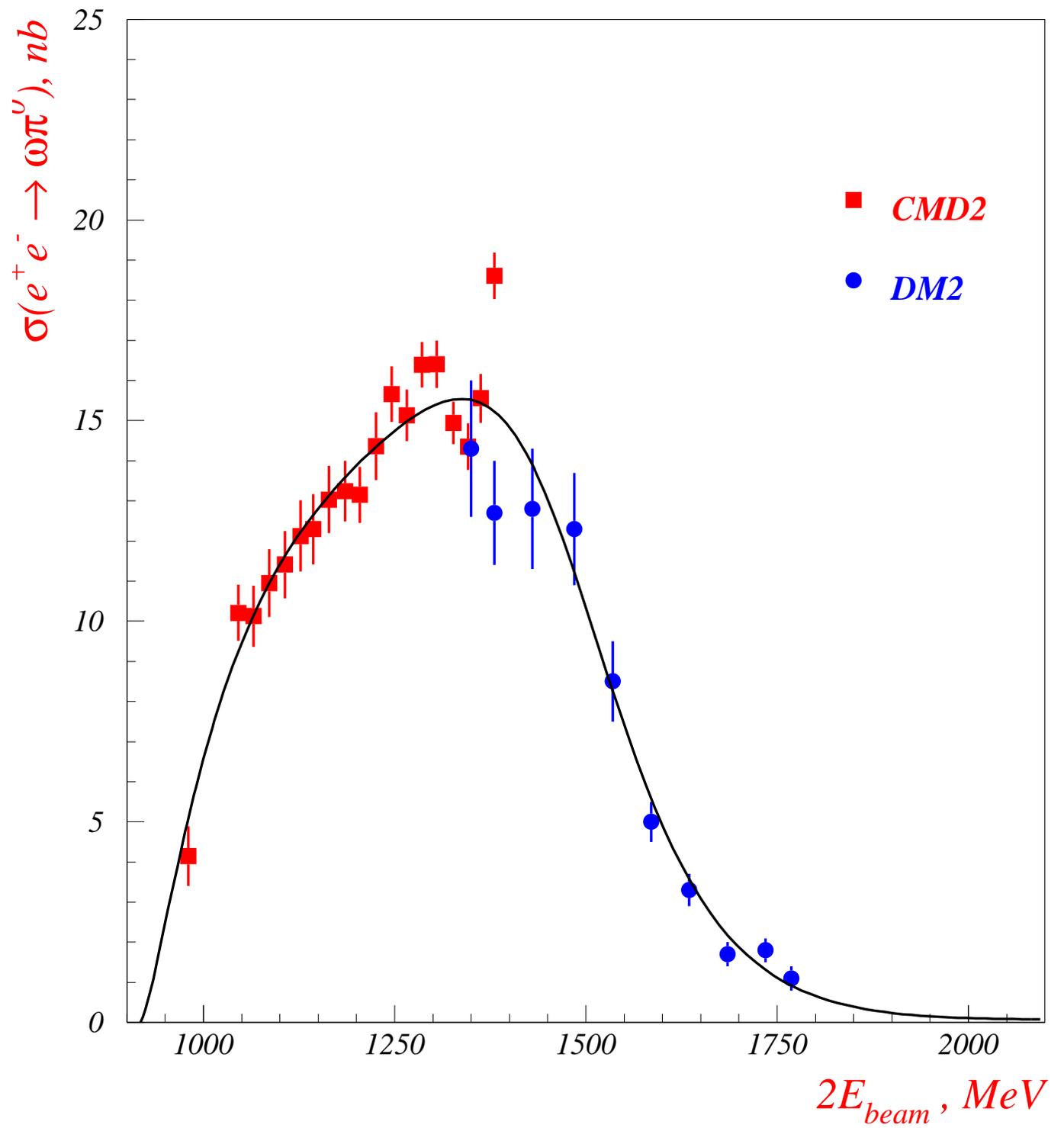,height=20cm}
\caption{
Energy dependence of the $e^+e^-\to \omega\pi^0$ cross section
}
\label{xsomega}
\end{figure}

\begin{figure}[ptb]
\centering
\epsfig{figure=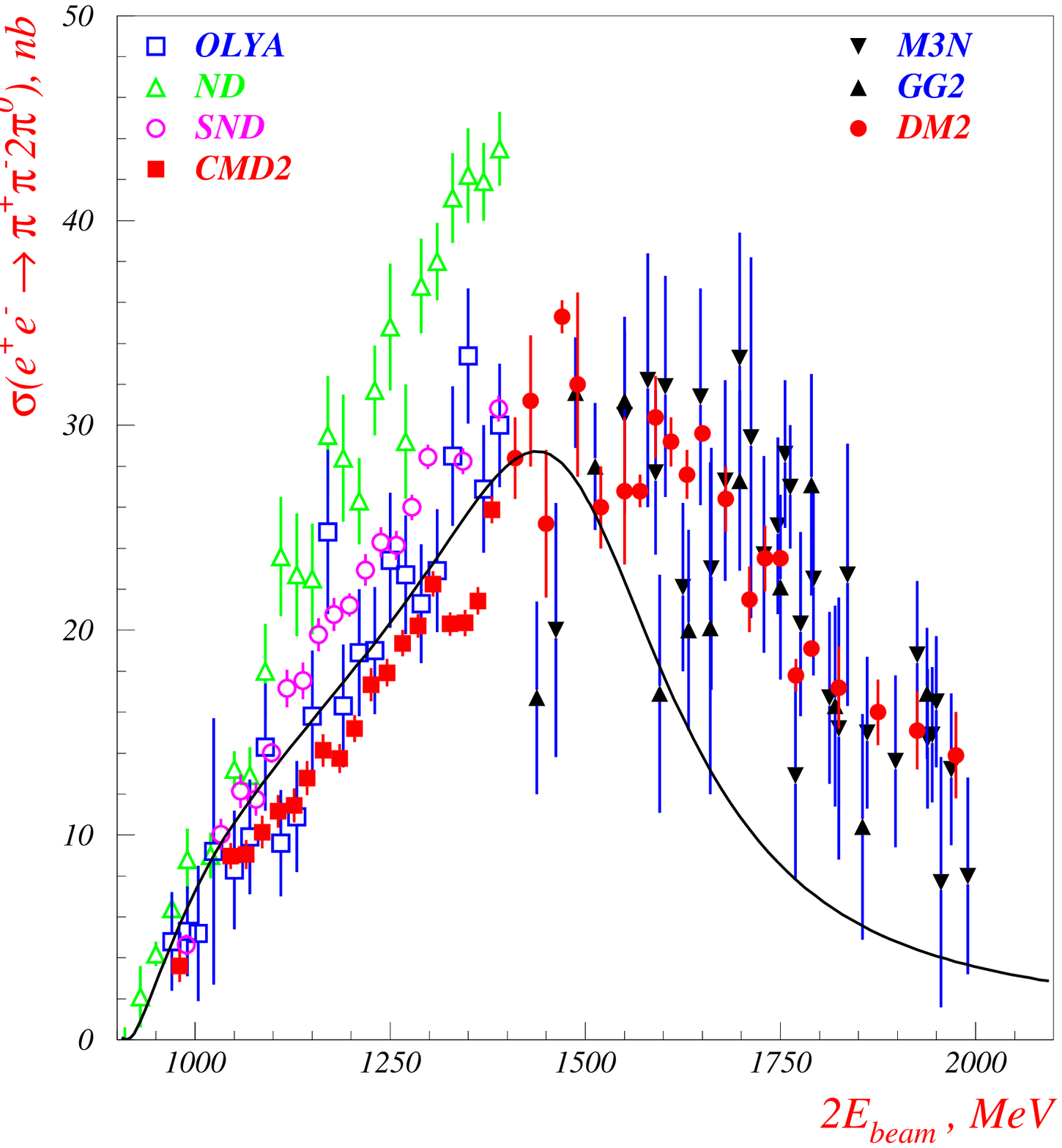,height=20cm}
\caption{
Energy dependence of the $e^+e^-\to \pi^+\pi^-2\pi^0$ cross section
}
\label{xs2tau}
\end{figure}

To get  $R_{2\pi^+2\pi^-}$ and
 $R_{\omega\pi}$ we used the cross section of $e^+e^-$ annihilation in the
energy range 1.--2. GeV obtained by several experimental groups and
represented in  Figs.~\ref{xs4tau} and ~\ref{xsomega}
 together with our fit. 
Our approximation of the experimental 
data took into account possible uncertainties in the data.
Assuming that the total cross section $e^+e^-\to 2\pi^+2\pi^-$ is saturated
by the $a_1\pi$ mechanism, we calculated the cross section
 $e^+e^-\to \pi^+\pi^-2\pi^0$.
The comparison of our approximation with
the experimental data is shown in Fig.~\ref{xs2tau}. 
One can see some difference
between our expectation and data at energies above 1.6 GeV. We
suppose that this difference can be explained by the possible systematic
errors or by contributions of another intermediate states which
become more important at higher 
energies. However, we expect that this difference does not affect very much 
our predictions for $\tau$  since this energy region has the
additional suppression factor due to limited phase space
in case of $\tau$ decay (see \eq{G1}). In order to check this
we compared the distribution over four-pion
invariant mass with the recent data obtained by CLEOII \cite{Cleo2} and
obtained that our interpolation of $e^+e^-$
data was in acceptable agreement with the $\tau$ decay data.

\begin{figure}[ptb]
\centering
\epsfig{figure=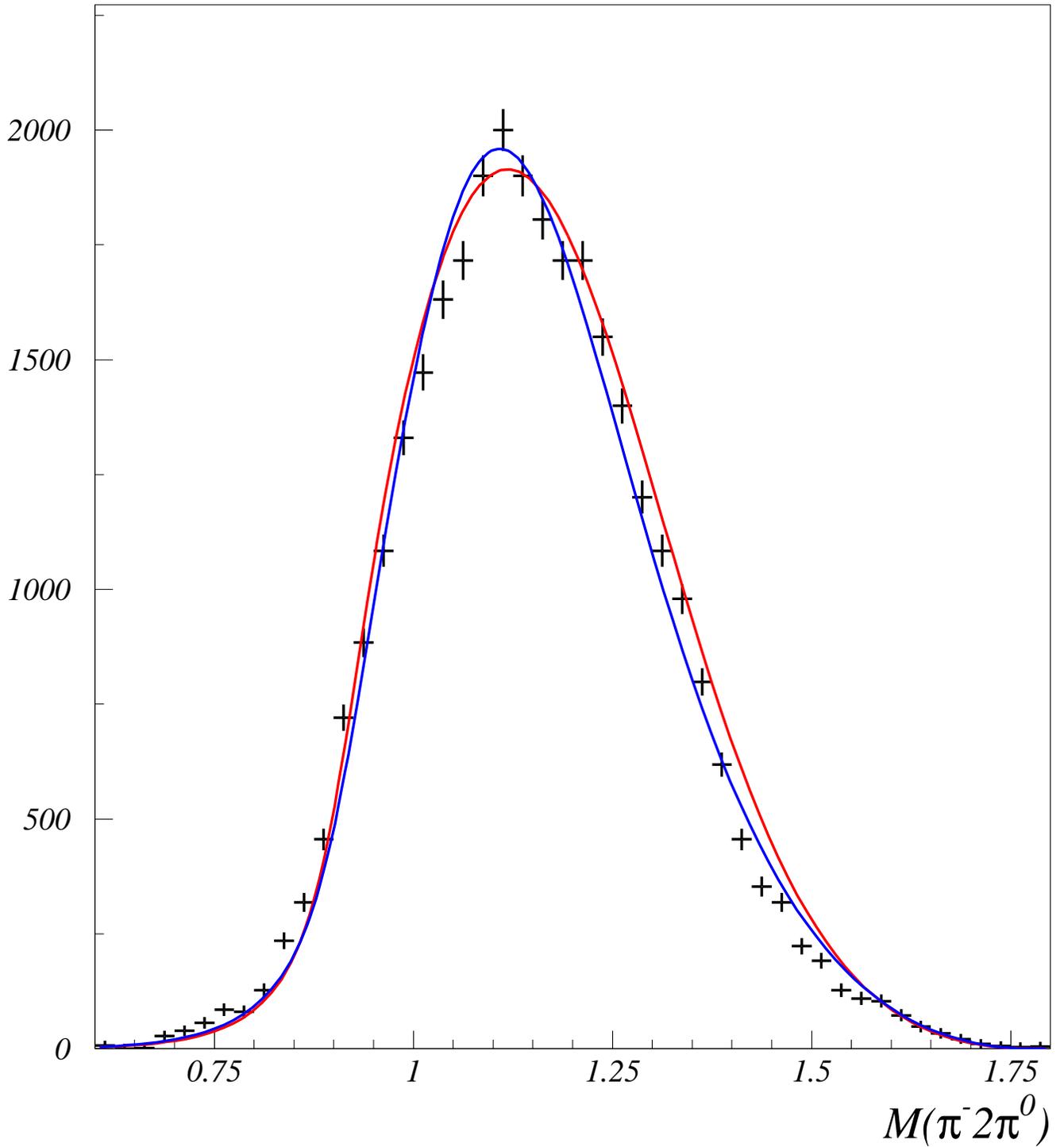,height=20cm}
\caption{
Three pion invariant mass distribution in the $\tau^-\to\pi^-2\pi^0\nu_\tau$
decay. Blue curve is obtained taking into account $a_1\to\sigma\pi$ decay.
Red one corresponds only to  $a_1\to\rho\pi$ decay. 
}
\label{am3pi}
\end{figure}

\begin{figure}[ptb]
\centering
\epsfig{figure=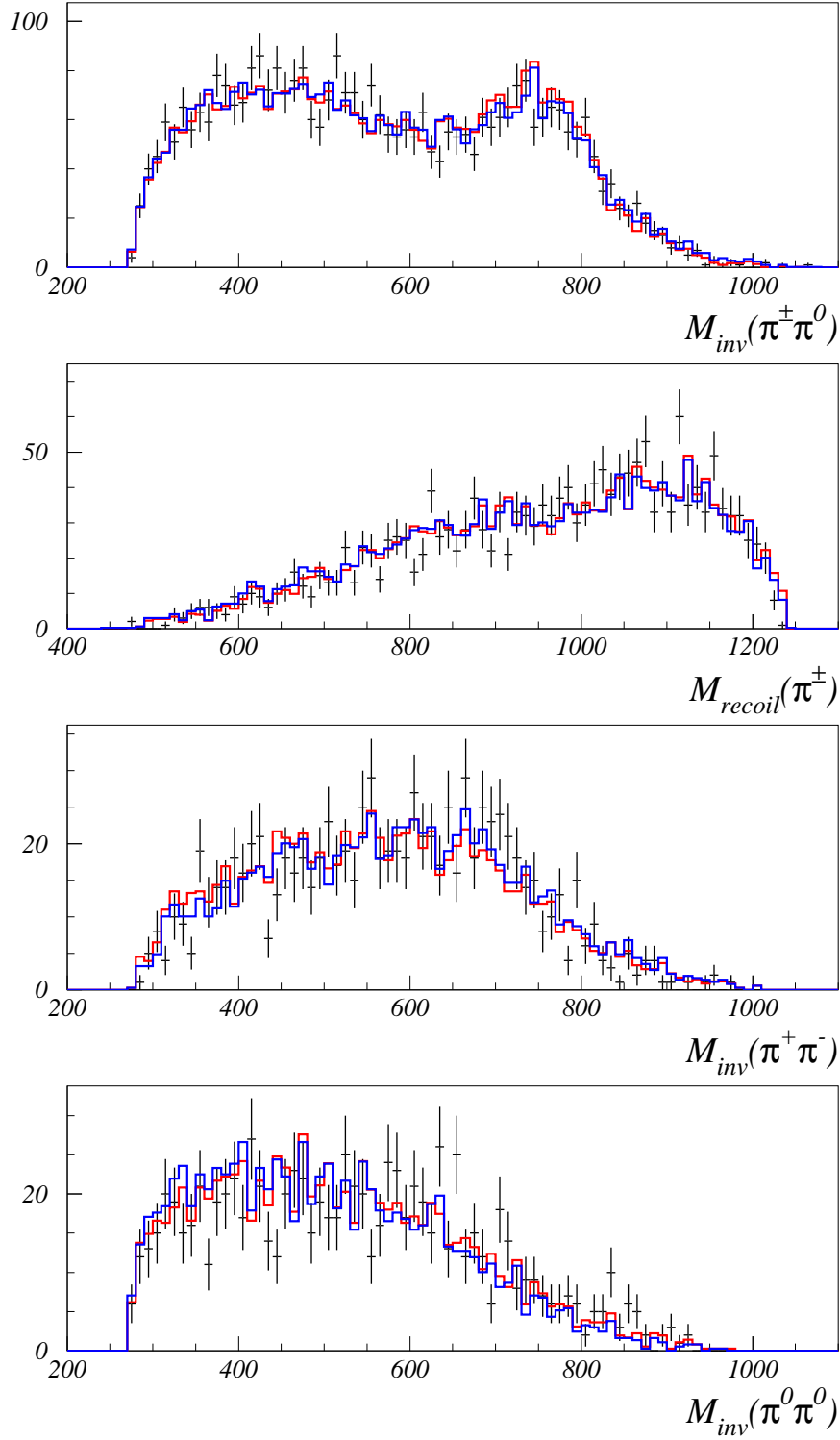,height=20cm}
\caption{
Distributions over invariant mass $M_{inv}(\pi^\pm\pi^0)$,
 $M_{recoil}(\pi^\pm)$,  $M_{inv}(\pi^+\pi^-)$,  $M_{inv}(\pi^0\pi^0)$
 for  $e^+e^-\to\pi^+\pi^-2\pi^0$ process after $\omega\pi^0$ events
subtraction.  Blue curve is obtained taking into account $a_1\to\sigma\pi$
decay. Red one corresponds only to  $a_1\to\rho\pi$ decay.
}
\label{slide1_a1_690}
\end{figure}

\begin{figure}[ptb]
\centering
\epsfig{figure=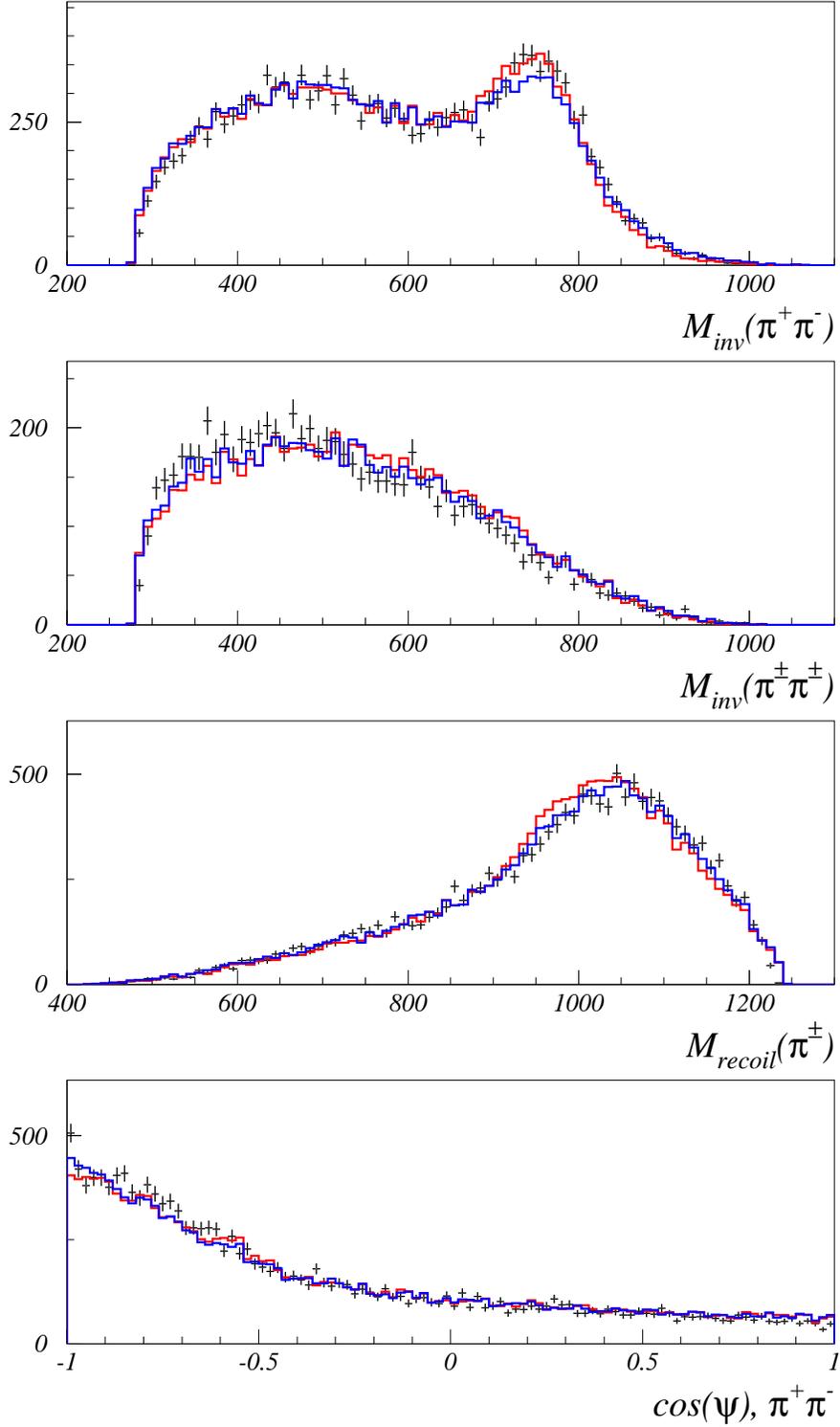,height=20cm}
\caption{
Distributions over invariant mass $M_{inv}(\pi^+\pi^-)$,
$M_{inv}(\pi^\pm\pi^\pm)$, $M_{recoil}(\pi^\pm)$, and
the distribution over the angle between the momenta of $\pi^+$ and $\pi^-$
 for  $e^+e^-\to 2\pi^+2\pi^-$ process.  Blue curve is obtained
taking into account $a_1\to\sigma\pi$ decay. Red one corresponds only
to  $a_1\to\rho\pi$ decay. 
}
\label{slide4_4piboth}
\end{figure}

In order to fix the parameters of our model we have used the data of
 $e^+e^-\to 4\pi$ \cite{our} and $\tau^-\to 2\pi^0\pi^-2\nu_\tau$
\cite{3pi}. Since the data in \cite{our} were obtained below the 
threshold of $a_1\pi$ creation it was difficult to extract independently
both the mass and width of $a_1$. The mass of $a_1$ was taken from
the PDG table \cite{PDG} and the width was obtained as a result of
optimal description of $e^+e^-\to \pi^+\pi^-2\pi^0$ data.
 Here we use the same $a_1$ mass and
the width we get as a result of optimal description by our matrix element
(see Appendix) of three pion
invariant mass distribution in $\tau^-\to 2\pi^0\pi^-\nu_\tau$ decay
\cite{3pi}. We used this way since the data of $\tau$ decay are more
sensitive to the parameters of $a_1$ meson. The comparison of our 
description of  three pion invariant mass distribution with the data is
presented in Fig.~\ref{am3pi}. 
We obtained the value of $a_1$ width which also provides
a good description of  $e^+e^-\to 4\pi$ data 
(see Figs.~\ref{slide1_a1_690},~\ref{slide4_4piboth}).
In \cite{3pi} it was obtained the evidence that  $a_1$ meson has significant
probability to decay into three pions through $\sigma\pi$
intermediate state. The data analysis of  $e^+e^-\to 4\pi$ 
also confirmed this statement. Therefore, here we take into account the
admixture of $\sigma\pi$ to the $a_1$ decay amplitude and the
parameters of this admixture we extracted from $e^+e^-\to 4\pi$ data.
For comparison we also present the results obtained without 
 $\sigma\pi$ contribution taken into account.

\begin{figure}[ptb]
\centering
\epsfig{figure=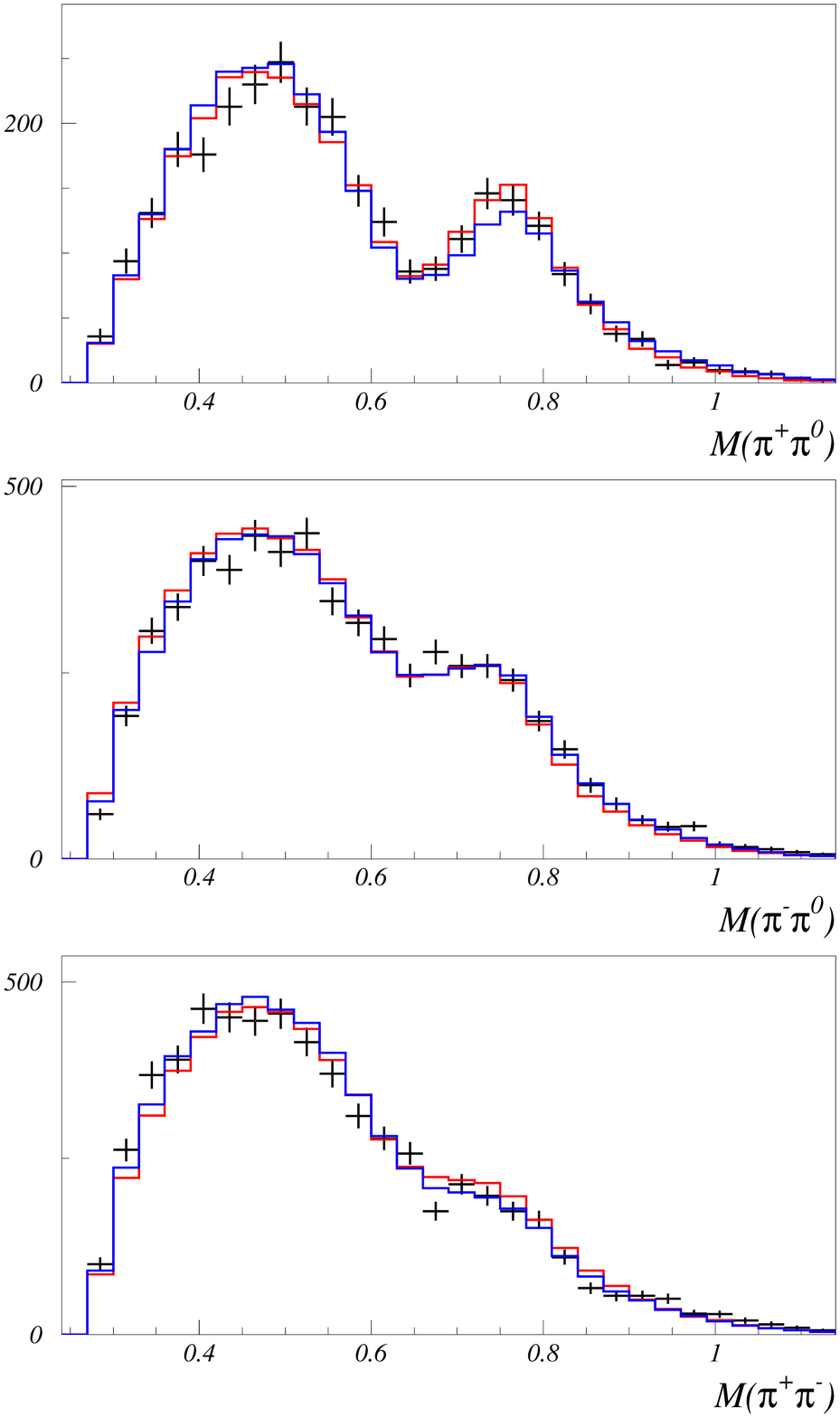,height=20cm}
\caption{
Distributions over invariant mass $M_{inv}(\pi^+\pi^0)$,
$M_{inv}(\pi^-\pi^0)$, and $M_{inv}(\pi^+\pi^-)$
for  $\tau^-\to 2\pi^-\pi^+\pi^0\nu_\tau$ decay obtained by CLEO \cite{Cleo1}.
Blue curve is the prediction obtained with $a_1\to\sigma\pi$ decay
taken into account. Red one corresponds only
to  $a_1\to\rho\pi$ decay. 
}
\label{cleo_all}
\end{figure}

\begin{figure}[ptb]
\centering
\epsfig{figure=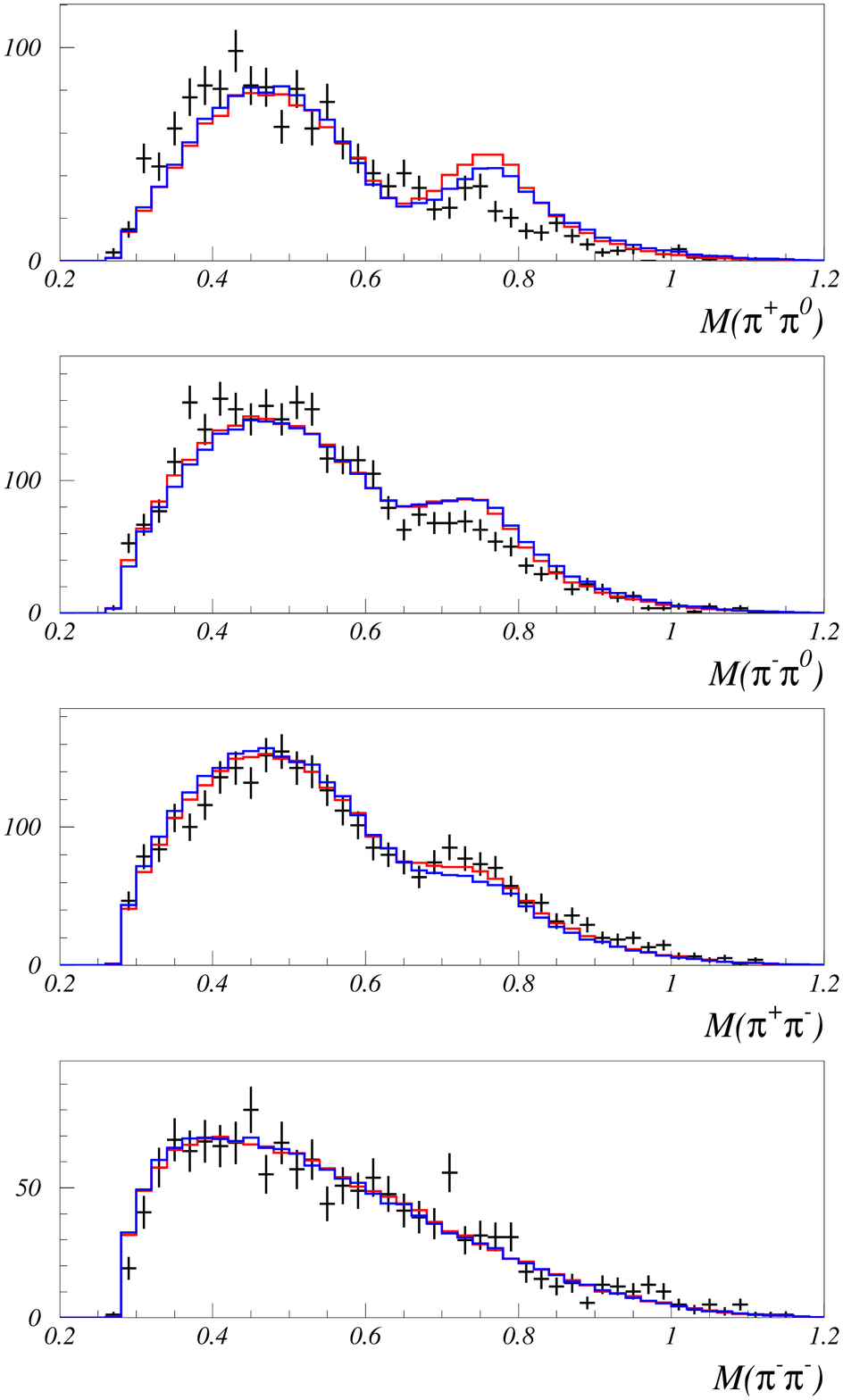,height=20cm}
\caption{
Distributions over invariant mass $M_{inv}(\pi^+\pi^0)$,
$M_{inv}(\pi^-\pi^0)$, $M_{inv}(\pi^+\pi^-)$, and $M_{inv}(\pi^-\pi^-)$
for  $\tau^-\to 2\pi^-\pi^+\pi^0\nu_\tau$ decay obtained by ARGUS
\cite{Argus}. 
Blue curve is the prediction obtained with $a_1\to\sigma\pi$ decay
taken into account. Red one corresponds only
to  $a_1\to\rho\pi$ decay. 
}
\label{argus_pic}
\end{figure}

When we get the parameters of our model, we can pass to the
description of  $\tau^-\to \pi^-\pi^+\pi^-\pi^0\nu_\tau$ decay. 
The most interesting information on the mechanism of four pion channel can be
obtained from two-pion mass distributions. 
Fig.~\ref{cleo_all} shows the comparison
of our predictions with the data of CLEOII detector \cite{Cleo1} obtained
without subtraction of $\omega\pi^-$ contribution. One can see a rather
good agreement with this data. The same comparison with the ARGUS data
\cite{Argus} is made in Fig.~\ref{argus_pic}. 
In this case our predictions are not
in a rough contradiction with the data, but the agreement is essentially
worse than with that of CLEOII, especially in the invariant mass range 
where the enhancement due to $\rho$ meson can be seen. 

\begin{figure}[ptb]
\centering
\epsfig{figure=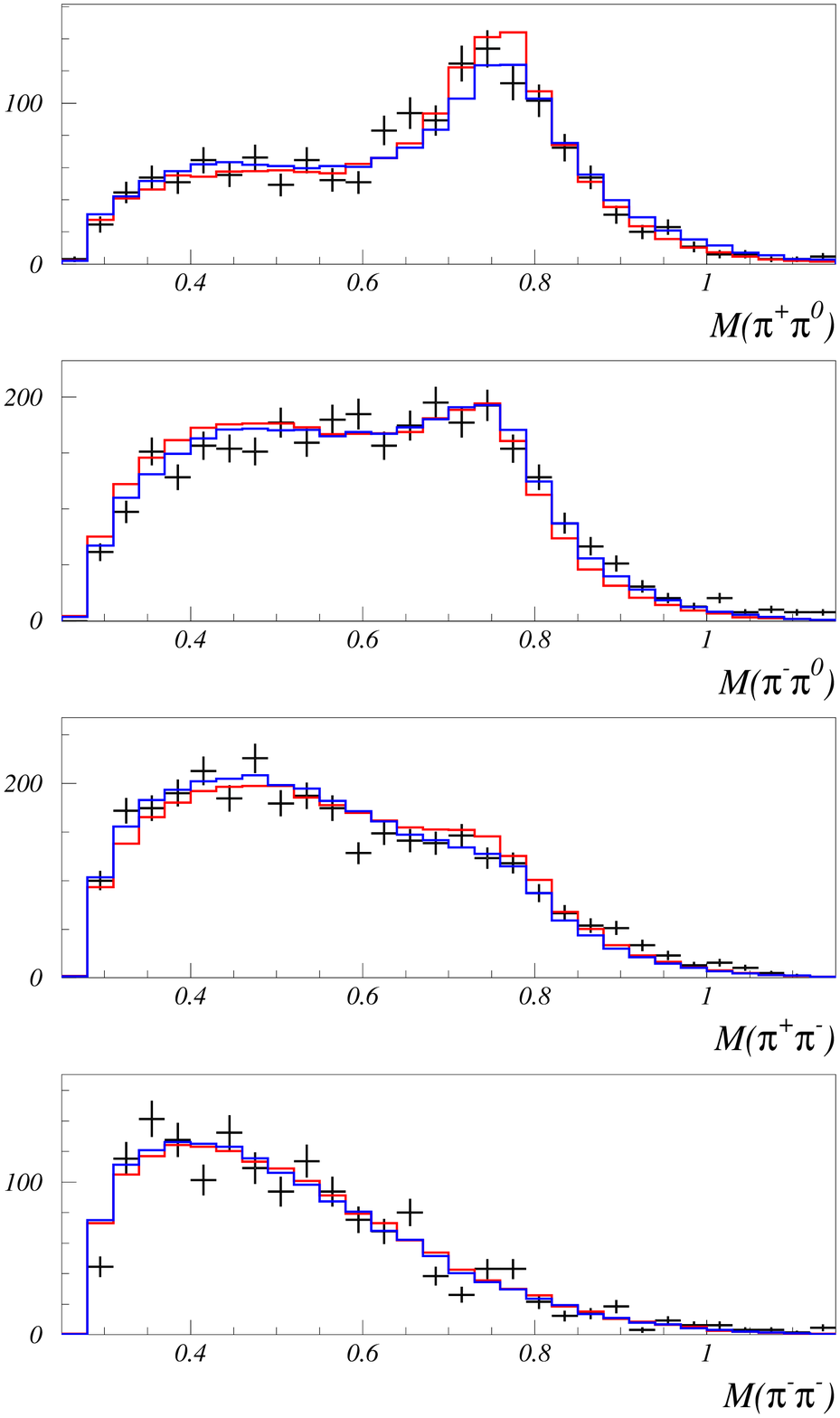,height=20cm}
\caption{
Distributions over invariant mass $M_{inv}(\pi^+\pi^0)$,
$M_{inv}(\pi^-\pi^0)$, $M_{inv}(\pi^+\pi^-)$, and $M_{inv}(\pi^-\pi^-)$
for  $\tau^-\to 2\pi^-\pi^+\pi^0\nu_\tau$ decay after $\omega\pi^-$
events subtraction obtained by ALEPH
\cite{Aleph}. Blue curve is the prediction obtained with
$a_1\to\sigma\pi$ decay taken into account. Red one corresponds only
to  $a_1\to\rho\pi$ decay. 
}
\label{aleph_pic}
\end{figure}

\begin{figure}[ptb]
\centering
\epsfig{figure=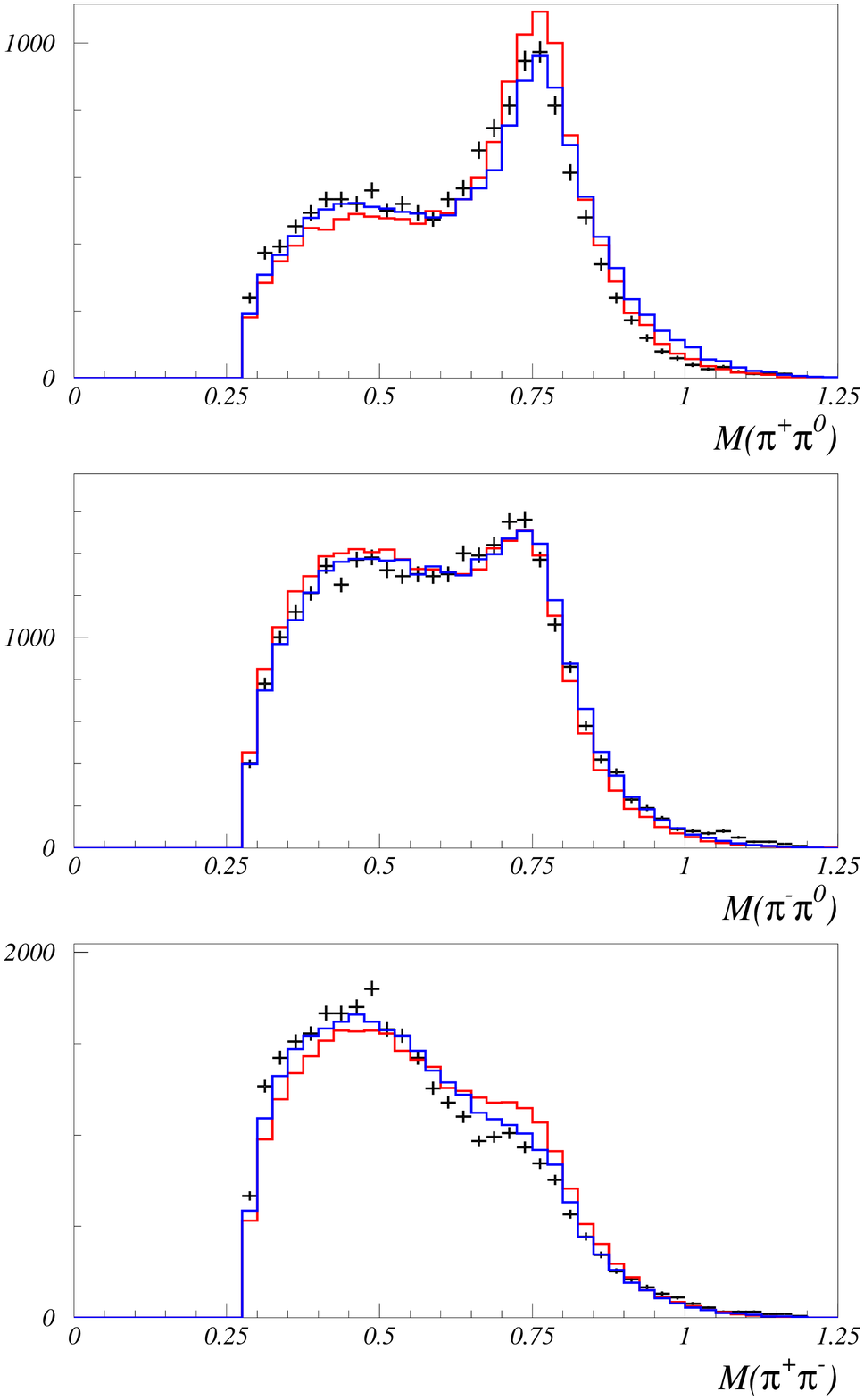,height=20cm}
\caption{
Distributions over invariant mass $M_{inv}(\pi^+\pi^0)$,
$M_{inv}(\pi^-\pi^0)$, and $M_{inv}(\pi^+\pi^-)$
for  $\tau^-\to 2\pi^-\pi^+\pi^0\nu_\tau$ decay after $\omega\pi^-$
events subtraction obtained by CLEO
\cite{Cleo2} . Blue curve is the prediction obtained with
$a_1\to\sigma\pi$ decay taken into account. Red one corresponds only
to  $a_1\to\rho\pi$ decay. 
}
\label{cleo_a1}
\end{figure}

The data obtained with the subtraction of $\omega\pi^-$ contribution allows
one to make a more detail comparison of the differential distributions
predicted within the assumption on $a_1\pi$ dominance. For this purpose
we have used the data obtained by ALEPH \cite{Aleph} 
(see Fig.~\ref{aleph_pic}) and
very recent high-statistics data of CLEOII \cite{Cleo2} 
(see Fig.~\ref{cleo_a1}).
In the first case we see a good agreement although we can not
take into account the detector efficiency and energy
resolution. The data of CLEOII, which have essentially higher statistics,
the agreement is a little bit worse. Unfortunately,  in this data
the contributions of background events (such as $K^0\pi^-\pi^0\nu_\tau$
and $K^{\pm}\pi^{\mp}\pi^-\pi^0\nu_\tau$ ) were not subtracted,
though their fraction was significant (about 8\%). The biggest
discrepancy can be seen for the $\pi^+\pi^-$ invariant mass distribution.
However the data of CLEOII in this case systematically differ from
data obtained by ALEPH (see Fig.~\ref{aleph_pic}c).

In spite of some disagreement between the data of different groups we
can conclude that the assumption of $a_1$ dominance
is in qualitative agreement with all available data. The quantitative
comparison of our model can be made only by full simulation which takes
into account the energy resolution and detector efficiency as well
as all possible background sources.
On the base of our analysis we come to conclusion that the previously
obtained upper limit (less than 44\% at 95\% CL) \cite{Argus}
for $a_1\pi\nu_\tau$ contribution to the $\tau\to 4\pi\nu_{\tau}$ decay
is hardly correct.
Note that in some theoretical works based on the effective chiral
theory of mesons \cite{Li} the predictions of $a_1\pi$ dominance were
made but without detail comparison with experimental data.

\section*{Acknowledgements}
The authors are grateful to all members of  
CMD-2 collaboration for useful discussions.

\section{Appendix: A model for $\epm\to 4\pi$ and $\tau\to
4\pi\nu_{\tau}$ processes}. 

To describe four pion production we use a simple model assuming
quasitwoparticle intermediate states and taking into account
the important effects of the identity of the final pions
as well as the interference of all possible amplitudes.
We use the notation  ${\tilde e}_{\mu}^a$
for the polarization vector of the initial hadron state ( $\tilde\rho$),
superscript corresponds to isospin. Here we consider the
contributions of  $a_1(1260)\pi$ and $\omega\pi)$  intermediate states
to the amplitudes $\tilde\rho\to 4\pi$ in the energy range under study.
Other contributions to these amplitudes from broad resonances
having the masses close to the threshold of $\rho\pi$ production
($a_2(1320)$, $\pi(1300)$ etc.) are discussed in \cite{our}.

As known, the form of the propagators $1/D(q)$ is very important for
analysing the data. It was found in \cite{our} that it is necessary
to take into account  the dependence of the imaginary part of the
propagators (width) on virtuality while
the corresponding corrections to the real part which can be expressed
through the imaginary part by dispersion relations are not
so important.
We represent the function $D(q)$ in the form
\beq
D(q)=q^2-M^2 +iM\Gamma \frac{g(q^2)}{g(M^2)} \, ,
\eeq
where $M$ and $\Gamma$  are the mass and width of the corresponding
particle, and the function $g(s)$ describes the dependence of the width
on the virtuality.
If $q^2=M^2$ then $D=iM\Gamma$,  in accordance with the usual
definition of  mass and  width of the resonance.
In the case of the $\rho$-meson we used the following representation for
the function $g_{\rho}(q)$:
\beq
g_{\rho}(s)=s^{-1/2}(s-4m^2)^{3/2}  \, ,
\eeq
where $m$ is the pion mass.

We include into consideration two the most important channels of
$a_1\to 3\pi$ decay : $a_1\to\rho\pi\to 3\pi$ and
$a_1\to\sigma\pi\to 3\pi$.
Taking into account the quantum numbers of pion and $a_1$-meson,
we can write the matrix elements corresponding to the processes
$\tilde\rho(P)\to a_1(q)\pi(p)$,
$a_1(q)\to\rho(P^{\prime})\pi(p)$, and
$a_1(q)\to\sigma(P^{\prime})\pi(p)$ as
\beqn\label{M1}
T(\tilde\rho\to a_1\pi)&=&F_{\tilde\rho a_1\pi}\eps^{abc}
(P_{\mu}{\tilde e}_{\nu}^a-P_{\nu}{\tilde e}_{\mu}^a)
q_{\mu}A_{\nu}^{b*}\phi^{c*}\, ,\\
T(a_1\to \rho\pi)&=&F_{a_1\rho\pi}\eps^{abc}
q_{\mu}A_{\nu}^{a}
(P_{\mu}^{\prime}e_{\nu}^{b*}-P_{\nu}^{\prime}e_{\mu}^{b*})\phi^{c*} \, ,
\nonumber \\
T(a_1\to \sigma\pi)&=&F_{a_1\sigma\pi}
(q_{\mu}A_{\nu}^{a}-q_{\nu}A_{\mu}^{a})
P_{\mu}^{\prime}p_{\nu}\phi^{a*} \, ,
\nonumber
\eeqn
where $a,b,c$ are isospin indices, $A_{\mu}^a$ and  $e_{\mu}^a$ are
the polarization vectors of $a_1$- and $\rho$-mesons, $\phi^b$ is the
pion wave function,
$F_{a_1\rho\pi}$ , $F_{a_1\sigma\pi}$ and
$F_{\tilde\rho a_1\pi}$ are form factors
depending on the virtuality of initial and final particles.
The explicit form of these form factors in the energy range considered
is not very essential (it contributes to the theoretical uncertainty
of the model). The matrix element of the transition
 $T(\rho\to \pi\pi)$ and  $T(\sigma\to \pi\pi)$  reads
\beqn\label{Mrho}
T(\rho\to \pi\pi)&=&F_{\rho\pi\pi}\eps^{abc}
e_{\mu}^{a}(p_{\mu}^{(b)}-p_{\mu}^{(c)})\phi^{b*}\phi^{c*}\, ,\nonumber\\
T(\sigma\to \pi\pi)&=&F_{\sigma\pi\pi}\phi^{a*}\phi^{a*}\, ,
\eeqn
where $p_{\mu}^{(b)}$ and $p_{\mu}^{(c)}$ are 4-momenta of the corresponding
pions. The matrix element $T(\tilde\rho\to 4\pi)$ in the rest frame of
$\tilde\rho$ can be written as $T=\tilde{\bf e}{\bf J}$ , where
$\tilde{\bf e}$ is the polarization vector of  $\tilde\rho$. The
corresponding probability of $\tilde\rho$ decay is given by the usual
formula
\beq\label{W}
dW(s)=\frac{|T|^2}{2E}\,(2\pi)^4\delta^{(4)}(\sum_{i=1}^4 p_i-P)
\,\prod_{i=1}^4\frac{d\p_i}{2\eps_i(2\pi)^3}\  ,
\eeq
where $P$ is the initial 4-momentum of $\tilde\rho$
( $P^0=E=\sqrt{s}$ , ${\bf P}=0$), $p_i=(\eps_i, \p_i)$
are the momenta of pions.
Note that due to the helicity conservation in $\epm$ annihilation
only transverse space components (with respect
to electron and positron momenta) of the 4-vector ${\tilde e}_{\mu}$ are
not zero.
Using  \eq{M1} and  \eq{Mrho} we obtain the following expressions for
the contributions of the $a_1(1260)$-meson to the current
$ {\bf J}_{a_1}= {\bf J}_{a_1\to\rho\pi}+{\bf J}_{a_1\to\sigma\pi}$ in
the process \\
$\tilde\rho^0\to \pi^+{(p_1)}\pi^+{(p_2)}\pi^-{(p_3)}\pi^-{(p_4)}$ :
\beqn\label{JAPPMM}
{\bf J}_{a_1\to\rho\pi}^{++--}\!\!\!&=&\!\!\!
G\left[\T_{1}(p_1,p_2,p_3,p_4)+\T_{1}(p_1,p_4,p_3,p_2)
+\T_{1}(p_2,p_1,p_3,p_4)
\right. \nonumber\\
&&+\T_{1}(p_2,p_4,p_3,p_1)+\T_{1}(p_1,p_2,p_4,p_3)
+\T_{1}(p_1,p_3,p_4,p_2) \nonumber \\
&&+\left.\T_{1}(p_2,p_1,p_4,p_3)
+\T_{1}(p_2,p_3,p_4,p_1)\right] \, ,  \\
{\bf J}_{a_1\to\sigma\pi}^{++--}\!\!\!&=&\!\!\!
G\left[\T_{2}(p_1,p_2,p_3,p_4)-\T_{2}(p_1,p_4,p_3,p_2)
+\T_{2}(p_2,p_1,p_3,p_4)
\right. \nonumber\\
&&-\T_{2}(p_2,p_4,p_3,p_1)+\T_{2}(p_1,p_2,p_4,p_3)
-\T_{2}(p_1,p_3,p_4,p_2) \nonumber \\
&&+\left.\T_{1}(p_2,p_1,p_4,p_3)
-\T_{2}(p_2,p_3,p_4,p_1)\right] \, , \nonumber
\eeqn
where
\beqn\label{j}
\dst
\T_{1}(p_1,p_2,p_3,p_4)=\frac{F_{a_1}^2(P-p_4)}
{D_{a_1}(P-p_4)D_{\rho}(p_1+p_3)}\times \\
\dst
\times\{(E-\eps_4)[\p_1(E\eps_3-p_4p_3)-\p_3(E\eps_1-p_4p_1)] \nonumber \\
\dst
- \p_4[\eps_1(p_4p_3)-\eps_3(p_4p_1)]\}\, ,
\nonumber\\
\dst\nonumber\\
\dst
\T_{2}(p_1,p_2,p_3,p_4)=\frac{z\,F_{a_1}^2(P-p_4)}
{D_{a_1}(P-p_4)D_{\sigma}(p_1+p_3)}\times \nonumber\\
\dst
\times(P-p_4)^2\,[(E-\eps_4)\p_2+\eps_2\p_4] \nonumber
\eeqn
Here $1/D_A(q)$ ,  $1/D_{\rho}(q)$ and  $1/D_{\sigma}(q)$
are propagators of $a_1$  , $\rho$ and $\sigma$  mesons, $F_{a_1}(q)$ is the
form factor, $G$ is some constant, 
$z$ is the  dimensionless complex constant.
Similarly to \eq{JAPPMM}, we obtain for the contributions of
$a_1(1260)$ to the current $ {\bf J}_{a1}$ in the process
$\tilde\rho^0\to \pi^+{(p_1)}\pi^-{(p_4)}\pi^0{(p_2)}\pi^0{(p_3)}$:
\beqn\label{JAPM00}
{\bf J}_{a_1\to\rho\pi}^{+-00}&=&G\left[-\T_{a_1}(p_4,p_2,p_3,p_1)
+\T_{a_1}(p_1,p_2,p_3,p_4)\right. \\
&&-\left.\T_{a_1}(p_4,p_3,p_2,p_1)+\T_{a_1}(p_1,p_3,p_2,p_4)\right]\, ,
\nonumber  \\
{\bf J}_{a_1\to\sigma\pi}^{+-00}\!\!\!&=&\!\!\!
G\left[\T_{2}(p_2,p_1,p_3,p_4)-\T_{2}(p_2,p_4,p_3,p_1)
\right. \nonumber \, .
\eeqn

The contributions of the $a_1(1260)$-meson to the currents  $ {\bf J}_{a_1}$ in
the process\\
$\tilde\rho^-\to \pi^+{(p_1)}\pi^0{(p_2)}\pi^-{(p_3)}\pi^-{(p_4)}$
reads :
\beqn\label{JAPOMM}
{\bf J}_{a_1\to\rho\pi}^{+0--}\!\!\!&=&\!\!\!
-G\left[\T_{1}(p_1,p_3,p_4,p_2)+\T_{1}(p_1,p_4,p_3,p_2)
+\T_{1}(p_2,p_1,p_3,p_4)
\right. \nonumber\\
&&+\left.\T_{1}(p_2,p_1,p_4,p_3)-\T_{1}(p_1,p_3,p_2,p_4)
-\T_{1}(p_1,p_4,p_2,p_3) \right] \, ,  \\
{\bf J}_{a_1\to\sigma\pi}^{+0--}\!\!\!&=&\!\!\!
G\left[\T_{2}(p_1,p_3,p_4,p_2)+\T_{2}(p_1,p_4,p_3,p_2)-
\right. \nonumber\\
&&\left.  -\T_{2}(p_1,p_2,p_3,p_4)-\T_{2}(p_1,p_2,p_4,p_3)
\right] \, , \nonumber
\eeqn

and in the process
$\tilde\rho^-\to \pi^0{(p_1)}\pi^0{(p_2)}\pi^0{(p_3)}\pi^-{(p_4)}$
it reads :
\beqn\label{JAOMMM}
{\bf J}_{a_1\to\rho\pi}^{0---}\!\!\!&=&\!\!\!
G\left[\T_{1}(p_4,p_1,p_2,p_3)+\T_{1}(p_4,p_1,p_3,p_2)
+\T_{1}(p_4,p_2,p_1,p_3)
\right. \nonumber\\
&&+\left.\T_{1}(p_4,p_2,p_3,p_1)+\T_{1}(p_4,p_3,p_1,p_2)
+\T_{1}(p_4,p_3,p_2,p_1) \right] \, ,  \\
{\bf J}_{a_1\to\sigma\pi}^{0---}\!\!\!&=&\!\!\!
G\left[\T_{2}(p_1,p_4,p_2,p_3)+\T_{2}(p_1,p_4,p_3,p_2)+
\T_{2}(p_2,p_4,p_3,p_1)-\right. \nonumber\\
&&\left.-\T_{2}(p_1,p_2,p_3,p_4)
-\T_{2}(p_1,p_3,p_2,p_4)-\T_{2}(p_2,p_3,p_1,p_4)
\right] \, . \nonumber \\
\eeqn

The function $g_{a_1}(s)$ in the propagator of $a_1$ has the form:
\beqn\label{g}
g_{a_1}(s)&=& F_{a_1}^2(Q) \int \left\{\left |
 \frac{\eps_2\p_1-\eps_1\p_2}{D_{\rho}(p_1+p_2)}+
 \frac{\eps_2\p_3-\eps_3\p_2}{D_{\rho}(p_2+p_3)}
+ \frac{z\sqrt{s}\p_2}{D_{\sigma}(p_1+p_3)}\right|^2+\right. \nonumber \\
&&+\left. \frac{|z|^2s}{3!}\left |
\frac{\p_1}{D_{\sigma}(p_2+p_3)}+
\frac{\p_2}{D_{\sigma}(p_1+p_3)}+\frac{\p_3}{D_{\sigma}(p_1+p_2)}\right|^2
\right\}\times \nonumber \\
&& \times \frac{d\p_1\,d\p_2\,d\p_3\,\delta^{(4)}(p_1+p_2+p_3-Q)}
 {2\eps_12\eps_2 2\eps_3(2\pi)^5}\, ,
\eeqn
where $Q^0=\sqrt s$ and ${\bf Q}=0$. The first term in the braces corresponds
to the decay $a_1\to\pi^+\pi^-\pi^0$, and the second one to $a_1\to 3\pi^0$.
The function $g_\sigma(s)$ in the propagator of $\sigma$ is equal to
\beq
g_{\sigma}(s)=(1-4m^2/s)^{1/2}  \, .
\eeq
 As a form factor, we used the function
$F(q)=(1+m_{a_1}^2/\Lambda^2)/(1+q^2/\Lambda^2)$
with $\Lambda\sim$ 1~GeV. We found that in the energy region under
discussion the amplitudes are not very sensitive to a value of $\Lambda$.

The amplitude of the process $\tilde\rho({\cal P})\to \omega(q)\pi(p)$
has the form:
\beq\label{om}
T(\tilde\rho\to \omega\pi)=F_{\tilde\rho\omega\pi}
\eps_{\mu\nu\alpha\beta}{\cal P}_{\mu}
q_{\nu}\tilde{e}_{\alpha}^a e_{\beta}^*\phi^{a*}\, ,
\eeq
where $ e_{\beta}$ is the polarization vector of the $\omega$-meson.
The matrix element of the transition $\omega\to\rho\pi$
can be written in the similar form. The  $\omega$-meson
contributes to  channels
$\tilde\rho^0\to \pi^+{(p_1)}\pi^0{(p_2)}\pi^0{(p_3)}\pi^-{(p_4)}$,
$\tilde\rho^-\to \pi^+{(p_1)}\pi^-{(p_2)}\pi^-{(p_3)}\pi^0{(p_4)}$ and
$\tilde\rho^+\to \pi^-{(p_1)}\pi^+{(p_2)}\pi^+{(p_3)}\pi^0{(p_4)}$
The corresponding current is equal to
\beqn\label{JOMPM00}
{\bf J}_{\omega}&=&
G_{\omega}\left[\T_{\omega}(p_2,p_4,p_1,p_3)-
\T_{\omega}(p_2,p_1,p_4,p_3)\right.\nonumber\\
&&-\left.\T_{\omega}(p_2,p_3,p_1,p_4)
\right] + (p_2\leftrightarrow p_3) \, ,
\eeqn
where
\beqn
\dst
\T_{\omega}(p_1,p_2,p_3,p_4)=
\frac{F_{\omega}^2(P-p_1)}{D_{\omega}(P-p_1)D_{\rho}(p_3+p_4)}\\
\dst
\times \{(\eps_4\p_3-\eps_3\p_4)(\p_1\p_2)-
\p_2(\eps_4\p_1\p_3-\eps_3\p_1\p_4) \nonumber \\
\dst
-\eps_2[\p_3(\p_1\p_4)-\p_4(\p_1\p_3)]\}
\, , \nonumber
\eeqn
$\eps_i$ is the energy of the corresponding pion, $F_{\omega}(q)$ is
the form factor.
Since the width of the $\omega$ is small,
we set $g_{\omega}(s)=1$
in the propagator $D_{\omega}(q)$.

\newpage

\end{document}